\DocumentMetadata{}
\documentclass[sigconf]{acmart}

\usepackage{multirow}
\usepackage{adjustbox}
\usepackage{tabularx}
\usepackage{booktabs}
\usepackage{array}
\usepackage{amsmath}
\usepackage{algorithm}
\usepackage{algorithmic}
\usepackage{graphicx}
\usepackage{textcomp}
\usepackage{xcolor}
\usepackage{comment}
\usepackage{newtxmath}
\usepackage{newtxtext,newtxmath}

\AtBeginDocument{%
  \providecommand\BibTeX{{%
    \normalfont B\kern-0.5em{\scshape i\kern-0.25em b}\kern-0.8em\TeX}}}

\setcopyright{acmcopyright}
\copyrightyear{2025}
\acmYear{2025}
\acmDOI{10.1145/xxxxxxx.xxxxxxx}

\acmConference[ACM]{21st ACM Conference on Computer and Communications Security}{}{} 

\acmPrice{15.00}
\acmISBN{978-1-4503-XXXX-X/21/06}

\begin{document}

\title[Hybrid Reputation Aggregation for Adversarial FL]{Hybrid Reputation Aggregation: A Robust Defense Mechanism for Adversarial Federated Learning in 5G and Edge Network Environments}

\author{Saeid Sheikhi}
\email{saeid.sheikhi@oulu.fi}
\orcid{0000-0002-3600-966X}
\affiliation{%
  \institution{Center for Ubiquitous Computing, University of Oulu}
  \city{Oulu}
   \country{Finland}
  \postcode{90570}
}

\author{Panos Kostakos}
\email{panos.kostakos@oulu.fi}
\orcid{0000-0002-8545-599X}
\affiliation{%
  \institution{Center for Ubiquitous Computing, University of Oulu}
  \city{Oulu}
  \country{Finland}
  }

\author{Lauri Loven}
\email{Lauri.Loven@oulu.fi}
\orcid{0000-0001-9475-4839}
\affiliation{%
  \institution{Center for Ubiquitous Computing, University of Oulu}
  \city{Oulu}
  \country{Finland}
  }
\renewcommand{\shortauthors}{Sheikhi et al.}

\begin{abstract}
Federated Learning (FL) in 5G and edge network environments face severe security threats from adversarial clients. Malicious participants can perform label flipping, inject backdoor triggers, or launch Sybil attacks to corrupt the global model. This paper introduces \emph{Hybrid Reputation Aggregation (HRA)}, a novel robust aggregation mechanism designed to defend against diverse adversarial behaviors in FL without prior knowledge of the attack type. HRA combines geometric anomaly detection with momentum-based reputation tracking of clients. In each round, it detects outlier model updates via distance-based geometric analysis while continuously updating a trust score for each client based on historical behavior. This hybrid approach enables adaptive filtering of suspicious updates and long-term penalization of unreliable clients, countering attacks ranging from backdoor insertions to random noise Byzantine failures. We evaluate HRA on a large-scale proprietary 5G network dataset (3M+ records) and the widely used NF-CSE-CIC-IDS2018 benchmark under diverse adversarial attack scenarios. Experimental results reveal that HRA achieves robust global model accuracy of up to 98.66\% on the 5G dataset and 96.60\% on NF-CSE-CIC-IDS2018, outperforming state-of-the-art aggregators such as Krum, Trimmed Mean, and Bulyan by significant margins. Our ablation studies further demonstrate that the full hybrid system achieves 98.66\% accuracy, while the anomaly-only and reputation-only variants drop to 84.77\% and 78.52\%, respectively, validating the synergistic value of our dual-mechanism approach. This demonstrates HRA's enhanced resilience and robustness in 5G/edge federated learning deployments, even under significant adversarial conditions.
\end{abstract}



\maketitle

\section{Introduction}
Federated learning (FL) has emerged as a cornerstone of distributed model training for edge devices and 5G networks, enabling collaborative learning without centralized data collection \cite{loghin2020disruptions,mcmahan2017communication}. By allowing a large number of devices (e.g., smartphones, IoT sensors, or edge nodes) to train a shared model locally and only send model updates to a central server, FL addresses privacy concerns and reduces communication costs \cite{sheikhi2023ddos}. However, the security of FL in these distributed settings is a growing concern. Adversaries can exploit the decentralized nature of FL to inject malicious model updates that degrade or manipulate the global model \cite{gong2023agramplifier,chen2025trustworthy}. In the context of 5G and edge computing, where deployments involve massive numbers of devices and potentially untrusted participants, such adversarial behavior poses a critical threat \cite{zhang2022fldetector,blika2024federated}.

\par \textbf{Adversarial Threats in FL:} A range of attacks have been demonstrated against federated learning. In \emph{data poisoning} or \emph{label-flipping} attacks, malicious clients intentionally mislabel their local training data or introduce corrupted samples, causing the global model to learn incorrect correlations. More insidious are \emph{backdoor attacks}, where an adversary embeds a hidden trigger in the model such that it performs normally on benign inputs but produces attacker-chosen outputs when the trigger is present \cite{bagdasaryan2020backdoor}. Other adversaries may launch \emph{Byzantine} attacks by sending arbitrary or noisy model updates that disrupt convergence \cite{blanchard2017machine}. A particularly powerful threat in large-scale networks is the \emph{Sybil attack}, in which one attacker controls multiple fake or compromised clients to wield disproportionate influence on the aggregation process \cite{fung2018mitigating}. These threats are exacerbated in 5G/edge environments due to the scale, heterogeneity, and dynamic membership of clients, compromised devices or Sybils can join the federation undetected and coordinate sophisticated attack strategies.

\par \textbf{Limitations of Existing Defenses:} To safeguard federated learning, researchers have proposed various robust aggregation rules that aim to tolerate a fraction of malicious clients. Notable examples include Krum \cite{blanchard2017machine}, Bulyan \cite{mhamdi2018hidden}, coordinate-wise median \cite{yin2018byzantine}, trimmed mean \cite{yin2018byzantine}, and the geometric median-based aggregator \cite{pillutla2022robust}. These algorithms use statistical techniques to identify and downweight outlier updates each round. For instance, Krum selects the update that is closest to others in the parameter space, excluding outliers, while Bulyan builds on Krum by iteratively filtering and averaging to improve Byzantine robustness. While effective against certain attacks, these defenses have significant limitations. Many require an estimate of the maximum number of adversarial clients ($f$) to tune their filtering thresholds, and their performance can degrade if the actual attack deviates from assumptions. Moreover, most existing aggregators treat each round independently, lacking a mechanism to \emph{learn} which clients are consistently unreliable. In non-IID settings typical of edge networks, even honest clients can occasionally produce divergent updates (e.g., due to unique local data), which a purely distance-based filter might mistakenly flag and remove. Conversely, clever adversaries can adapt their updates to evade one-shot outlier detection by mimicking normal client behavior. Thus, static, per-round defenses struggle to adapt to evolving or stealthy attack patterns in a long-running federated training session

In this paper, we argue that a more \emph{dynamic and holistic} approach to aggregation is needed for adversarial FL in 5G and edge scenarios. Our key insight is to combine instantaneous anomaly detection with historical behavior tracking, to differentiate between one-off benign outliers and truly malicious actors. We propose a novel aggregation strategy called \textbf{Hybrid Reputation Aggregation (HRA)} that integrates geometric anomaly detection with momentum-based reputation scoring. At a high level, HRA works as follows: in each training round, the server analyzes the submitted model updates in the high-dimensional model space and computes an anomaly score for each update based on its distance from the majority (using a geometric approach akin to outlier detection). Simultaneously, HRA maintains a \emph{reputation score} for each client, which is updated over time (with a momentum factor) depending on whether the client's updates have been identified as suspicious or benign in previous rounds. Clients whose updates consistently deviate from the consensus will experience a decline in reputation, reducing their influence on the aggregated model in future rounds. On the other hand, clients with a history of reliable contributions retain higher weights. This hybrid mechanism allows HRA to rapidly filter obviously malicious updates in each round \emph{and} adapt to persistent attack sources over multiple rounds.

HRA is designed to be \emph{attack-agnostic}: it does not rely on any specific signature of a particular attack type, nor does it require knowing how many clients might be compromised. Instead, it dynamically builds trust scores and flags anomalies based on observed behavior. This makes it robust against a wide spectrum of attacks, including unforeseen or adaptive ones, without tuning parameters for each scenario. We implement HRA in a standard FL framework and evaluate it under a variety of adversarial conditions. Our experiments involve a proprietary 5G network dataset containing over 3 million data records, which simulates a realistic edge federated learning scenario with non-IID data across hundreds of clients. We test HRA against strong attackers employing Sybil strategies (multiple colluding adversaries), targeted model poisoning (label flips and backdoors), and untargeted random-noise attacks. 

Experimental evaluations demonstrate that HRA substantially outperforms traditional robust aggregation methods in both detecting malicious updates and preserving global model accuracy. In federated learning simulations, HRA achieved a test accuracy of \textbf{98.66\%} on our proprietary 5G testbed data and maintained \textbf{96.60\%} accuracy on the NF-CSE-CIC-IDS2018 benchmark, while classical defenses achieved significantly lower accuracies. For instance, on the 5G dataset, Krum only achieved 23.73\%, Trimmed Mean 22.85\%, and Median 71.24\% accuracy under the same adversarial conditions. Our synergy ablation study further validates the hybrid approach, showing that the full HRA system achieves 98.66\% accuracy while the anomaly-only variant drops to 84.77\% and the reputation-only variant to 78.52\%. Moreover, due to its efficient integration of geometric anomaly detection with momentum-based reputation tracking, HRA introduces only a negligible computational overhead at the server, making it highly suitable for deployment in latency-sensitive 5G and edge network environments.

\par \textbf{Contributions:} In summary, this paper makes the following contributions:
\begin{itemize}
    \item We propose \textbf{Hybrid Reputation Aggregation (HRA)}, a new robust aggregation mechanism for federated learning that combines geometric anomaly detection with momentum-based client reputation tracking. HRA is a general defense method that operates without prior knowledge of attack patterns and adapts to various adversarial behaviors (Sybil attacks, poisoning, backdoors, etc.) on the fly.
    \item We design HRA specifically for challenging 5G/edge network FL environments, featuring dynamic client populations and non-IID data. The reputation system in HRA provides a memory of client behavior, enabling the aggregator to distinguish transient outliers from persistent adversaries, which improves reliability for honest clients and the detection of malicious ones.
    \item We perform extensive experiments on both a proprietary 5G network dataset with over 3~million data records, simulating a large-scale edge FL deployment, and the publicly available NF-CSE-CIC-IDS2018 benchmark. We evaluate HRA against multiple attack types (label-flipping data poisoning, backdoor insertion, noise injection, and Sybil collusion) and compare it with state-of-the-art robust aggregators like Krum, Bulyan, median, and trimmed mean. HRA consistently achieves higher accuracy and robustness; for example, it retains up to \textbf{98.66\%} test accuracy on the 5G dataset and \textbf{96.60\%} on NF-CSE-CIC-IDS2018, representing substantial improvements over existing defenses. The following best method, Bulyan, achieves only 96.15\% on the 5G dataset and 88.73\% on NF-CSE-CIC-IDS2018.
\end{itemize}

\section{Related Work}
\noindent \textbf{Adversarial Attacks on Federated Learning.} The vulnerability of FL to adversarial clients has been well-documented in recent studies. \emph{Poisoning attacks} entail malicious clients altering their local training data or labels to corrupt the global model’s performance. Bhagoji et al.~\cite{bhagoji2019analyzing} analyze model poisoning attacks in FL and show that even a small fraction of corrupted clients can significantly impact model accuracy. In a simple label-flipping scenario, attackers might, for instance, flip the labels of examples from one class to another, causing the global model to misclassify that class. More targeted are \emph{backdoor attacks}, where the adversary’s goal is to make the model respond incorrectly to inputs containing a specific trigger while behaving normally otherwise. Bagdasaryan et al.~\cite{bagdasaryan2020backdoor} demonstrated that a single determined attacker can implement a backdoor in an FL setting by modifying its model update (and possibly scaling it) in one training round, achieving a targeted misclassification (e.g., making the global model classify images with a certain sticker as a different category). Another class of attacks is \emph{Byzantine attacks}, in which adversaries send arbitrarily malformed updates, these could be random noise or carefully crafted vectors, to derail the training process. Blanchard et al.’s work on Krum \cite{blanchard2017machine} was motivated by such Byzantine faults, showing that without defenses, a few bad updates can prevent convergence or drastically lower accuracy. In addition, \emph{Sybil attacks} have been identified as a severe threat in federated networks: an attacker creates many pseudonymous clients to join the federation, amplifying its influence. Fung et al.’s FoolsGold defense \cite{fung2018mitigating} specifically examined this scenario, highlighting how multiple fake clients can cooperate to skew the learning process or evade detection. These studies underscore the importance of robust aggregation mechanisms that can handle both malicious outliers and collusion by adversaries.

\par \noindent \textbf{Robust Aggregation Techniques.} A number of robust aggregation rules have been proposed to defend FL against adversarial or unreliable clients. One of the earliest is \textbf{Krum} \cite{blanchard2017machine}, which chooses a single local model update that is closest to its $n-f-2$ nearest neighbors (where $f$ is the assumed maximum number of Byzantine clients) and uses it as the aggregated update. The idea is to exclude updates that are far from the majority under the assumption that honest updates will cluster together. \textbf{Multi-Krum} repeatedly applies Krum to select multiple updates (instead of one) before averaging to retain more information from honest clients. Building on this, \textbf{Bulyan} \cite{mhamdi2018hidden} is a two-step aggregator: first, it uses an approach like Multi-Krum to pick a set of candidate updates deemed likely to be benign; second, it computes the coordinate-wise trimmed mean of those candidates, further eliminating outlier values. Bulyan offers improved robustness guarantees, reducing the impact of Byzantine values to a smaller bound. Other approaches dispense with picking specific candidates and instead use robust statistics across all updates: the \textbf{coordinate-wise median} \cite{yin2018byzantine} takes the median of each model parameter over all client updates (which can tolerate up to 50\% arbitrary corruption on each dimension), and the \textbf{trimmed mean} \cite{yin2018byzantine} discards a certain fraction of the highest and lowest values for each parameter before averaging the rest. These methods assume that the majority of clients are honest for each parameter update. More recently, the \textbf{geometric median} has been explored for FL aggregation \cite{pillutla2022robust}, where the server finds a model update that minimizes the sum of distances to all client updates (this can be seen as a multivariate generalization of the median). The geometric median-based aggregator (sometimes called RFA for robust federated averaging) can offer strong theoretical robustness but typically requires an iterative solution and can be computationally heavier. 

While effective to varying degrees, all the above aggregation rules are inherently memoryless they operate based on the current round’s updates without considering the past behavior of clients. They also generally rely on knowing or assuming an upper bound on the fraction of adversaries ($f$). If $f$ is underestimated, these methods may fail to filter out all malicious updates; if $f$ is overestimated, they may throw away too many genuine updates, hurting model accuracy. Additionally, sophisticated attackers can sometimes bypass these defenses. For example, recent work by Fang et al.~\cite{fang2020local} showed that an adaptive group of attackers can carefully craft their updates to appear mutually consistent and relatively benign (thus avoiding detection by Krum or trimmed mean), yet still introduce a significant error into the global model.

\par \noindent \textbf{Reputation and Trust-Based Approaches.} An alternative line of defense in FL is to incorporate notions of trust or reputation into the aggregation process. Instead of treating each update in isolation, these approaches maintain the state of clients. \textbf{FoolsGold} \cite{fung2018mitigating}, for instance, does not use a traditional robust aggregator but dynamically adjusts the effective learning rate of each client based on the similarity of its updates to others. The intuition is that Sybil attackers will produce updates that are more similar to each other (since they have a common objective), so if a client’s update frequently aligns with others, FoolsGold penalizes it by reducing its influence. This implicitly builds a reputation in that clients performing unique (and hence likely honest) contributions are trusted more over time. \textbf{FLTrust} \cite{cao2020fltrust} takes a different approach by assuming the server has a small trustworthy dataset: it evaluates each client’s model update on this clean data to compute a trust score and uses those scores to weight the aggregation (clients that perform poorly on the validation data are likely malicious and get lower weight). Some recent frameworks have also suggested explicitly tracking a reputation score for each client across rounds, updating scores based on behaviors like update magnitude or consistency with the global model \cite{wang2022flare}. Such reputation-based methods can improve resilience, especially against repeating offenders, but they often require careful design to avoid new vulnerabilities (e.g., an attacker might behave well to gain trust and then attack later or collude to upvote each other if peer grading is used). Moreover, methods like FLTrust require additional trusted data, which may not be available in practice.

Therefore, prior work provides valuable techniques for robust aggregation and attack mitigation in federated learning, yet no single solution fully addresses the challenges of adversarial FL in a dynamic edge network setting. Our proposed HRA method differentiates itself by unifying instantaneous anomaly detection with long-term reputation tracking without relying on external trust data or assumptions of attacker prevalence. As we will show, this hybrid approach offers robust protection against a broad array of attacks, filling an important gap in the federated learning security landscape.

\section{Methodology}
\label{sec:methodology}

In this section, we describe our approach to defending FL systems against adversarial client behaviors. Our proposed framework integrates geometric anomaly detection and reputation-based weighting to mitigate the effects of malicious updates. First, we provide an overview of the FL setup and data pre-processing. Next, we detail the local training procedure with adversarial attack simulation, present the key aggregation algorithm with pseudocode and relevant equations, and finally describe the simulation setup, execution, and evaluation methodology.

\subsection{Federated Learning Setup and Data Preprocessing}

Let $\mathcal{D} = \{(\mathbf{x}_i, y_i)\}_{i=1}^{N}$ denote the complete training dataset, where $\mathbf{x}_i \in \mathbb{R}^d$ is the feature vector and $y_i \in \{0,1\}$ is the binary label (with $0$ indicating benign and $1$ an attack). Data preprocessing is performed as follows:
\begin{itemize}
    \item \textbf{Conversion:} All categorical and hexadecimal features are converted to numeric values.
    \item \textbf{Imputation:} Missing values are replaced by the median of each feature.
    \item \textbf{Normalization:} Each feature is standardized using
    \begin{equation}
      \mathbf{x}' = \frac{\mathbf{x} - \mu}{\sigma},
      \label{eq:normalization}
    \end{equation}
    where $\mu$ and $\sigma$ are the mean and standard deviation computed from the training data.
\end{itemize}

The training dataset is then partitioned uniformly among $M$ clients (e.g., $M=10$). For each client $j$, let $\mathcal{D}_j$ denote its local dataset.

\subsection{Local Training with Adversarial Attack Simulation}

Each client trains a local logistic regression model. The model output is computed as:
\begin{equation}
    \hat{y} = \sigma(z) = \frac{1}{1 + e^{-z}}, \quad \text{with } z = \mathbf{x}^\top \mathbf{w} + b.
    \label{eq:logistic}
\end{equation}
The training minimizes the binary cross-entropy loss via gradient descent. To simulate adversarial behavior, clients are assigned specific attack types (e.g., \texttt{label\_flipping}, \texttt{noise}, \texttt{backdoor}, \texttt{sybil}, and \texttt{sign\_flipping}). For instance, under a label-flipping attack, a client transforms its labels according to:
\begin{equation}
    y_{\text{adv}} = 1 - y.
    \label{eq:label_flipping}
\end{equation}

The key steps of local training with attack simulation can be summarized in pseudocode \ref{alg:local_training} as follows:

\begin{algorithm}[H]
\caption{Local Training with Attack Simulation}
\label{alg:local_training}
\begin{algorithmic}[1]
\REQUIRE Local data $\mathcal{D}_j$, global parameters $(\mathbf{w}, b)$, learning rate $\eta$, attack type $\mathcal{A}$, local epochs $E$
\STATE Initialize $\mathbf{w}_j \leftarrow \mathbf{w}$, $b_j \leftarrow b$
\IF {$\mathcal{A} =$ \texttt{label\_flipping}}
    \STATE Set $y \leftarrow 1 - y$ for all local samples
\ENDIF
\FOR{$e = 1$ to $E$}
    \STATE Compute predictions: $z \leftarrow X_j \mathbf{w}_j + b_j$
    \STATE Compute error: $\mathbf{e} \leftarrow \sigma(z) - y_j$
    \STATE Update weights: $\mathbf{w}_j \leftarrow \mathbf{w}_j - \eta \cdot \nabla_{\mathbf{w}}$, where $\nabla_{\mathbf{w}} = \frac{1}{|\mathcal{D}_j|} X_j^\top \mathbf{e}$
    \STATE Update bias: $b_j \leftarrow b_j - \eta \cdot \nabla_b$, where $\nabla_b = \frac{1}{|\mathcal{D}_j|}\sum \mathbf{e}$
\ENDFOR
\IF {$\mathcal{A} =$ \texttt{noise}}
    \STATE Perturb $\mathbf{w}_j$ and $b_j$ with Gaussian noise
\ELSIF {$\mathcal{A} =$ \texttt{sign\_flipping}}
    \STATE Invert the sign of the update and amplify it
\ELSIF {$\mathcal{A} =$ \texttt{backdoor}}
    \STATE Add a predefined trigger vector to $\mathbf{w}_j$
\ELSIF {$\mathcal{A} =$ \texttt{sybil}}
    \STATE Add large-scale perturbations to $\mathbf{w}_j$ and $b_j$
\ENDIF
\RETURN $(\mathbf{w}_j, b_j)$
\end{algorithmic}
\end{algorithm}

\subsection{Hybrid Reputation Aggregation}
After local updates, the server aggregates the client models. In addition to standard methods (mean, median, etc.), our HRA method is designed to weigh each client’s update based on its \emph{anomaly} and \emph{reputation}.

\subsubsection{Geometric Reference and Anomaly Scoring}

First, a robust reference update is computed via the geometric median of the client updates:
\begin{equation}
    \mathbf{w}_{\text{ref}} = \operatorname{GeomMed}\left(\{\mathbf{w}_j\}_{j=1}^{M}\right).
\end{equation}
Each client’s anomaly score is calculated as:
\begin{equation}
    \Delta_j = \|\mathbf{w}_j - \mathbf{w}_{\text{ref}}\|_2.
\end{equation}

\subsubsection{Reputation-Based Weighting}
Given two thresholds $T_{low}$ and $T_{high}$, the weight factor $\phi(\Delta_j)$ for client $j$ is defined as:
\begin{equation}
    \phi(\Delta_j) =
    \begin{cases}
        1, & \Delta_j \leq T_{low}, \\
        \frac{T_{high} - \Delta_j}{T_{high} - T_{low}}, & T_{low} < \Delta_j < T_{high}, \\
        0, & \Delta_j \geq T_{high}.
    \end{cases}
    \label{eq:weight_func}
\end{equation}
Additionally, each client is assigned a reputation score $r_j$ updated iteratively using a momentum parameter $\rho \in [0, 1]$:
\begin{equation}
    r_j^{(t+1)} = \rho \, r_j^{(t)} + (1-\rho) \, \phi(\Delta_j).
\end{equation}

\subsubsection{Aggregation Rule}
The final aggregated model is computed as a weighted average:
\begin{align}
    \mathbf{w}_{\text{agg}} &= \frac{\sum_{j=1}^{M} r_j\, \phi(\Delta_j)\, \mathbf{w}_j}{\sum_{j=1}^{M} r_j\, \phi(\Delta_j)}, \label{eq:agg_w} \\
    b_{\text{agg}} &= \frac{\sum_{j=1}^{M} r_j\, \phi(\Delta_j)\, b_j}{\sum_{j=1}^{M} r_j\, \phi(\Delta_j)}. \label{eq:agg_b}
\end{align}

The overall HRA procedure is summarized in the following pseudocode \ref{alg:hra}:

\begin{algorithm}[H]
\caption{Hybrid Reputation Aggregation}
\label{alg:hra}
\begin{algorithmic}[1]
\REQUIRE Client updates $\{(\mathbf{w}_j, b_j)\}_{j=1}^{M}$, reputations $\{r_j\}_{j=1}^{M}$, thresholds $T_{low}, T_{high}$
\STATE Compute $\mathbf{w}_{\text{ref}} = \operatorname{GeomMed}\left(\{\mathbf{w}_j\}\right)$
\FOR{each client $j$}
    \STATE $\Delta_j \leftarrow \|\mathbf{w}_j - \mathbf{w}_{\text{ref}}\|_2$
    \STATE Compute $\phi(\Delta_j)$ using Equation~\eqref{eq:weight_func}
\ENDFOR
\STATE Compute aggregated weights using Equations~\eqref{eq:agg_w} and~\eqref{eq:agg_b}
\RETURN $(\mathbf{w}_{\text{agg}}, b_{\text{agg}})$
\end{algorithmic}
\end{algorithm}

\subsection{Simulation Setup and Evaluation Methodology}
\label{sec:evaluation}
We evaluated our approach through extensive simulations under various adversarial conditions. The simulation process follows these steps:

\begin{enumerate}
    \item \textbf{Model Initialization:} A global logistic regression model with parameters $(\mathbf{w}, b)$ is initialized to zero.
    \item \textbf{Data Partitioning:} The preprocessed training dataset is partitioned into $M$ non-IID subsets for clients.
    \item \textbf{Local Training:} In each communication round $r$, each client performs local training (Algorithm~\ref{alg:local_training}) for a fixed number of epochs using its local data and possible adversarial modifications.
    \item \textbf{Aggregation:} The server aggregates client updates using HRA (Algorithm~\ref{alg:hra}) to update the global model.
    \item \textbf{Learning Rate Decay:} The learning rate is decayed as:
    \begin{equation}
         \eta_r = \eta_0 \cdot \gamma^r,
    \end{equation}
    where $\gamma < 1$ is the decay factor.
    \item \textbf{Evaluation:} At the end of each round, the updated global model is evaluated on a held-out test set. Evaluation metrics include accuracy, precision, recall, and F1 score. In addition, we monitor the average anomaly distance and reputation evolution.
    \item \textbf{Statistical Analysis:} To ensure the robustness of our results, simulations are repeated for multiple independent runs (e.g., $n=5$), and paired t-tests are performed for comparative analysis against baseline aggregation methods.
\end{enumerate}

All simulation parameters, such as the number of rounds $R$, client count $M$, learning rate $\eta_0$, decay rate $\gamma$, and thresholds $(T_{low}, T_{high})$, are chosen based on preliminary experiments.

\begin{figure*}[h]
\centering
\includegraphics[width=1.0\textwidth]{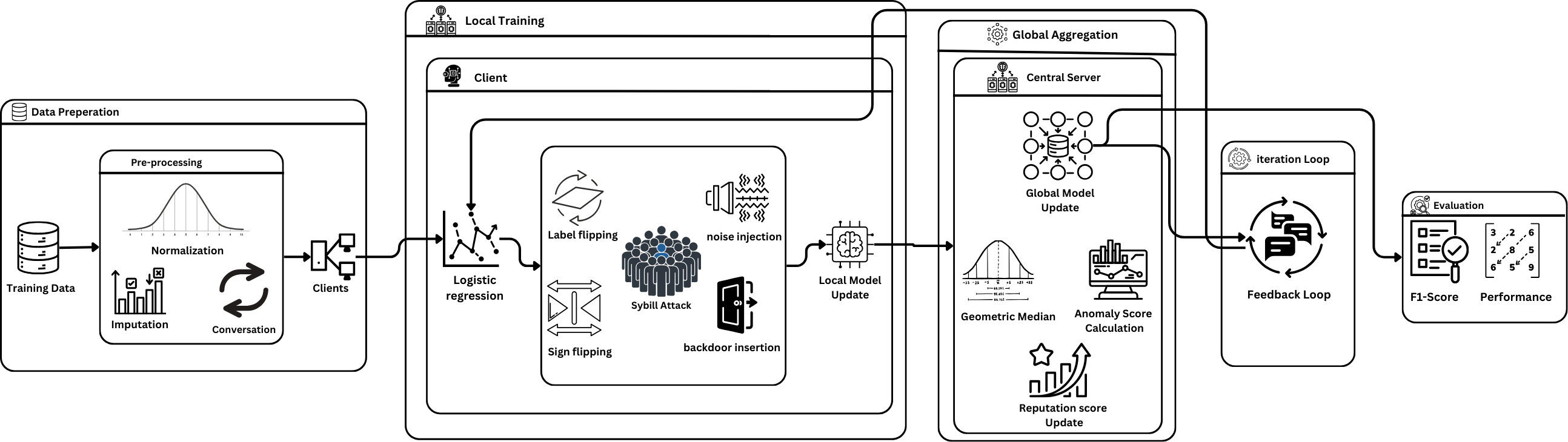}
\caption{Overall Architecture: From Data Preprocessing to Hybrid Reputation-Based Aggregation.}
\label{fig:framework}
\end{figure*}

The methodology integrates a comprehensive FL training process with adversarial attack simulation and a novel hybrid reputation aggregation scheme. By combining geometric anomaly detection and momentum-based reputation updating, our approach effectively discounts malicious client updates and maintains robust global model performance. Extensive simulations on both proprietary and public datasets (including large-scale 5G data) substantiate the superior performance of our method under a variety of adversarial scenarios.

\section{Experimental Results and Ablation Study}
\label{sec:results}

In this section, we evaluate the performance of our proposed hybrid reputation aggregation mechanism against several baseline aggregation methods in a federated learning setting under adversarial conditions. We first describe our experimental setup, including details of our dataset collection, attack simulations, and preprocessing. We then present performance comparisons on both a custom 5G dataset and the NF-CSE-CIC-IDS2018 \cite{sarhan2021netflow} benchmark. Finally, we report an ablation study on key HRA parameters and provide statistical validation via paired t-tests.

\subsection{Experimental Setup, Datasets, and Preprocessing}
\subsection{Datasets and Preprocessing}

We conduct our experiments on two datasets: a custom 5G network traffic dataset collected via a dedicated testbed, and the publicly available NF-CSE-CIC-IDS2018 \cite{sarhan2021netflow} benchmark. Table~\ref{tab:dataset_stats} summarizes the key dataset statistics after preprocessing, while Table~\ref{tab:class_distribution_5g} details the class distributions for the 5G dataset.

\begin{table}[ht]
  \centering
  \caption{Dataset Statistics after Preprocessing}
  \label{tab:dataset_stats}
  \begin{tabular}{lcc}
    \toprule
    \textbf{Statistic} & \textbf{5G Testbed} & \textbf{NF-CSE-CIC-IDS2018} \\
    \midrule
    Features           & 29         & 10         \\
    Train Samples      & 1,753,454  & 6,713,920  \\
    Test Samples       & 194,829    & 1,678,481  \\
    Total Samples      & 1,948,283  & 8,392,401  \\
    \bottomrule
  \end{tabular}
\end{table}

\begin{figure*}[ht]
    \centering
    \includegraphics[width=0.7\linewidth]{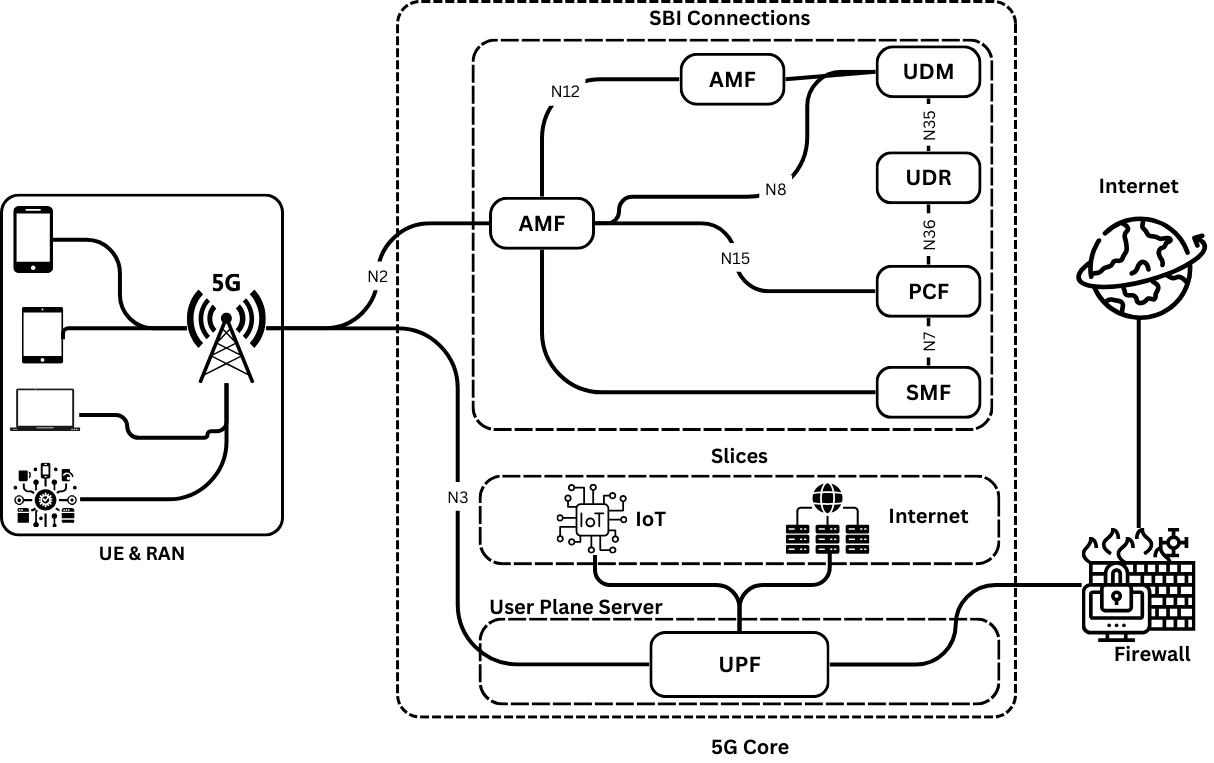}
    \caption{The structure of the 5G Core in the testbed.}
    \label{fig:5gcore}
\end{figure*}

\paragraph{5G Dataset Collection.}
The 5G dataset was captured in a private testbed environment built around an Open5GS core network with Dockerized services. The testbed supports both Internet and IoT traffic via network slicing. Figure \ref{fig:5gcore} presents the network design of the 5G testbed. A structured workflow was utilized to create a robust and realistic dataset that includes both normal and malicious traffic. The traffic collection involved:
\begin{enumerate}
    \item \textbf{Network Slicing:} Configuring different slices to isolate various traffic streams.
    \item \textbf{Service Simulation:} Deploying Dockerized services to emulate realistic network applications.
    \item \textbf{Traffic Generation:} Simulating both benign traffic and adversarial activities (including DoS\_MQTT, DDoS, Eavesdropping, MITM, SQL Injection, Unauthorized Data Access, Brute Force, and Device Spoofing).
    \item \textbf{Data Capture and Processing:} Continuously capturing traffic and extracting features followed by conversion, imputation, constant-feature removal, and standard normalization.
\end{enumerate}

The 5G dataset, prior to preprocessing, contains 29 features, with a training set of 1,753,454 samples and a testing set of 194,829 samples. For the conducted experiments, the class labels in the the  dataset were converted to a binary format. The original class distribution of the collected 5G dataset is shown in Table~\ref{tab:class_distribution_5g}.

\begin{table}[ht]
    \centering
    \caption{Class Distribution for the 5G Dataset (prior-Preprocessing)}
    \label{tab:class_distribution_5g}
    \begin{tabular}{lcc}
        \toprule
        \textbf{Attack Class} & \textbf{Train records} & \textbf{Test records} \\
        \midrule
        Benign                      & 66,631   & 7,404 \\
        DoS\_MQTT                   & 25,052   & 2,720 \\
        DDoS                        & 16,484   & 1,901 \\
        Eavesdropping               & 361      & 37    \\
        MITM                        & 68       & 11    \\
        SQL Injection               & 54       & 10    \\
        Unauthorized Data Access    & 31       & 10    \\
        Brute Force                 & 26       & 10    \\
        Device Spoofing             & 10       & 10    \\
        \bottomrule
    \end{tabular}
\end{table}

To ensure a strict evaluation of our approach, we processed and partitioned our datasets using a pre-processing procedure before federated learning simulations. The following outlines the procedures for each dataset.

\paragraph{NF-CSE-CIC-IDS2018 Dataset.}  
We employ the NF-CSE-CIC-IDS2018 dataset, a benchmark widely used for network intrusion detection. After initial data cleaning and removal of constant-valued features, we retain 10 critical features. The dataset is partitioned into a training set with 6,713,920 samples and a testing set with 1,678,481 samples. The preprocessing steps are as follows:
\begin{itemize}
    \item \textbf{Conversion:} Categorical and hexadecimal feature values are converted into numeric representations.
    \item \textbf{Imputation:} Missing values are imputed using the median of each feature.
    \item \textbf{Normalization:} Standard scaling is applied to each feature:
    \begin{equation}
      \mathbf{x}' = \frac{\mathbf{x} - \mu}{\sigma},
      \label{eq:normalization}
    \end{equation}
    where $\mu$ and $\sigma$ denote the mean and standard deviation computed from the training set.
\end{itemize}
For federated simulations, the training set is partitioned uniformly among 10 clients, ensuring that each client receives a statistically representative subset of the overall data.

\paragraph{Federated Data Partitioning.}
For all federated simulations, the respective training set is partitioned among 10 clients. To simulate a realistic non-IID environment, which is a key challenge in edge networks, a Dirichlet distribution ($\alpha=0.5$) was used to distribute the data labels unevenly across clients. This ensures that each client possesses a skewed and unique class distribution, rigorously testing the aggregator's performance under heterogeneous data conditions.

\subsection{Attack Simulation and Implementation Details}
During each FL round, clients perform local training using a logistic regression model. Adversarial behavior is simulated by assigning specific attack types to a subset of clients. For example, under a \texttt{label\_flipping} attack, local labels are inverted:
\begin{equation}
  y_{\text{adv}} = 1 - y.
  \label{eq:label_flipping_again}
\end{equation}
Other attack types (e.g., \texttt{noise}, \texttt{backdoor}, \texttt{sybil}, \texttt{sign\_flipping}) introduce controlled perturbations to the local model updates.

Our local training procedure (see Algorithm~\ref{alg:local_training}) runs for $E=16$ epochs per round with a decaying learning rate:
\[
  \eta_r = \eta_0 \cdot \gamma^r,
\]
where $\eta_0=0.1$ and $\gamma=0.998$. For the 5G dataset, our implementation uses all 23 features; for NF-CSE-CIC-IDS2018, we use the 10 selected features.

The experiments are implemented in Python. All simulation code (data partitioning, local training with adversarial attack simulation, and aggregation) is modular, allowing extensive reproducibility.

\subsection{Performance Comparison}

Tables~\ref{tab:results_5g} and~\ref{tab:results_cic} summarize the classification performance (Accuracy, Precision, Recall, F1 Score, and ROC AUC) of HRA alongside several baseline aggregation methods.

\begin{table}[h]
  \centering
  \renewcommand{\arraystretch}{1.4}
  \caption{Classification Performance on the 5G Dataset}
  \label{tab:results_5g}
  \resizebox{\linewidth}{!}{%
    \begin{tabular}{lccccc}
      \toprule
      \textbf{Method} & \textbf{Accuracy} & \textbf{Precision} & \textbf{Recall} & \textbf{F1 Score} & \textbf{ROC AUC} \\
      \midrule
      Hybrid Reputation & 0.9866 ± 0.0000 & 0.94999 & 0.98514 & 0.96724 & 0.98438 \\
      Bulyan           & 0.9615 ± 0.0000 & 0.94649 & 0.94457 & 0.94553 & 0.96385 \\
      FLTrust          & 0.7402 ± 0.2422 & - & - & - & - \\
      Median           & 0.7124 ± 0.1684 & 0.56289 & 0.54246 & 0.55249 & 0.70472 \\
      FLARE            & 0.6057 ± 0.2155 & - & - & - & - \\
      BaFFLe           & 0.5976 ± 0.2580 & - & - & - & - \\
      GeoMed           & 0.5331 ± 0.3517 & 0.82097 & 0.24536 & 0.37781 & 0.61423 \\
      Simple Mean      & 0.4731 ± 0.1367 & 0.22485 & 0.86634 & 0.35703 & 0.46159 \\
      Krum             & 0.2373 ± 0.1810 & 0.08645 & 0.27531 & 0.13158 & 0.17829 \\
      Trimmed Mean     & 0.2285 ± 0.0479 & 0.92242 & 0.50275 & 0.65079 & 0.74470 \\
      \bottomrule
    \end{tabular}
  }
\end{table}

\begin{table}[h]
  \centering
  \renewcommand{\arraystretch}{1.4}
  \caption{Classification Performance on the NF-CSE-CIC-IDS2018 Dataset}
  \label{tab:results_cic}
  \resizebox{\linewidth}{!}{%
    \begin{tabular}{lccccc}
      \toprule
      \textbf{Method} & \textbf{Accuracy} & \textbf{Precision} & \textbf{Recall} & \textbf{F1 Score} & \textbf{ROC AUC} \\
      \midrule
      Hybrid Reputation & 0.9660 ± 0.0000 & 0.82304 & 0.91763 & 0.86776 & 0.94518 \\
      Bulyan           & 0.8873 ± 0.0000 & 0.60199 & 0.28971 & 0.39117 & 0.63161 \\
      GeoMed           & 0.8675 ± 0.0154 & 0.05805 & 0.00873 & 0.01517 & 0.49458 \\
      FLTrust          & 0.8071 ± 0.1059 & - & - & - & - \\
      FLARE            & 0.6608 ± 0.2101 & - & - & - & - \\
      Simple Mean      & 0.6511 ± 0.1722 & 0.01288 & 0.03519 & 0.01886 & 0.33121 \\
      Median           & 0.5715 ± 0.0723 & 0.13341 & 0.55934 & 0.21543 & 0.52854 \\
      BaFFLe           & 0.5401 ± 0.0325 & - & - & - & - \\
      Trimmed Mean     & 0.5206 ± 0.0741 & 0.11416 & 0.33858 & 0.17075 & 0.48795 \\
      Krum             & 0.4910 ± 0.0520 & 0.12123 & 0.99789 & 0.21619 & 0.49899 \\
      \bottomrule
    \end{tabular}
  }
\end{table}

\subsubsection{Performance Evolution Plots}

\begin{figure}[t]
  \centering
  \includegraphics[width=0.98\linewidth]{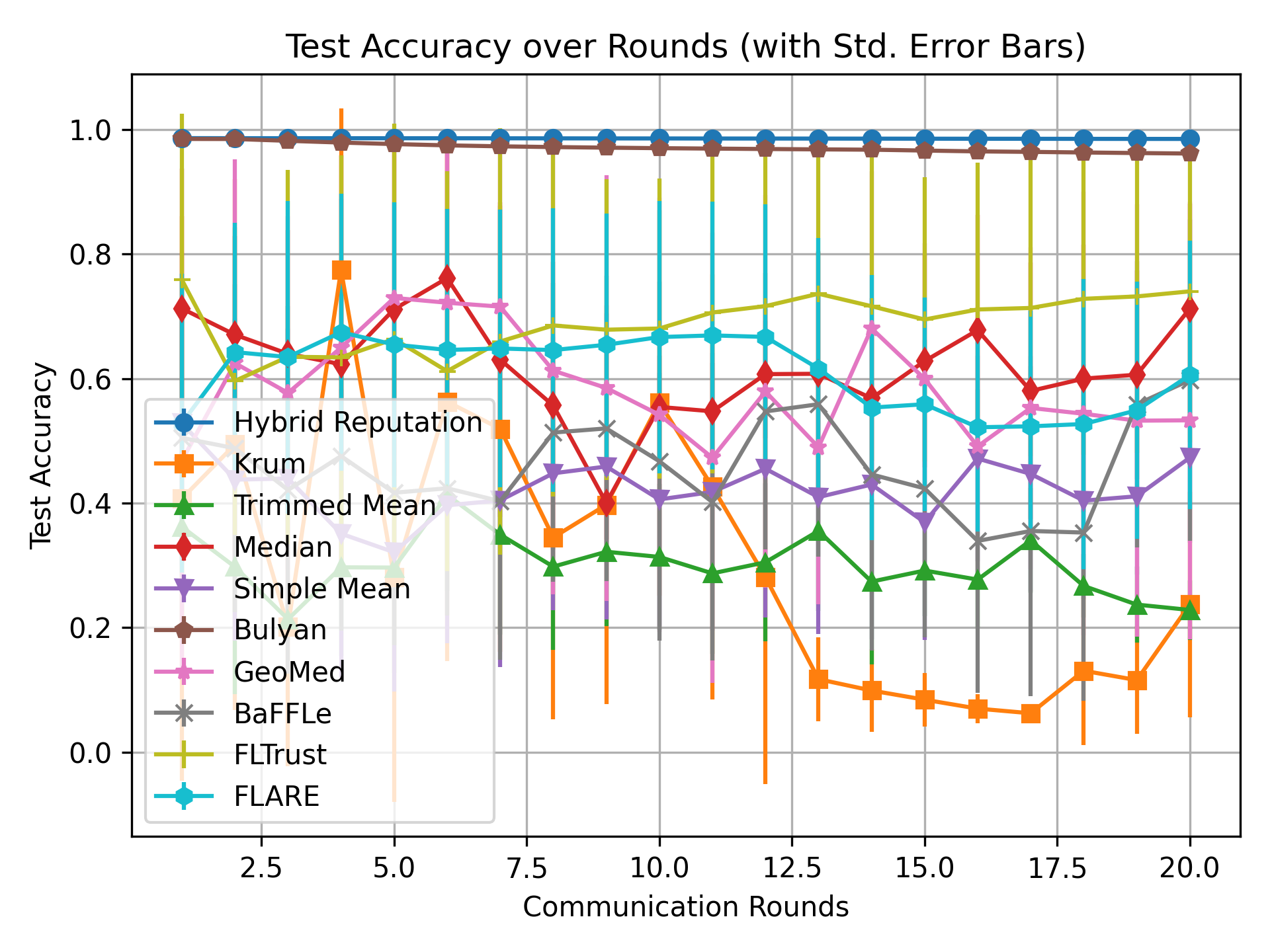}\\[1ex]
  \includegraphics[width=0.98\linewidth]{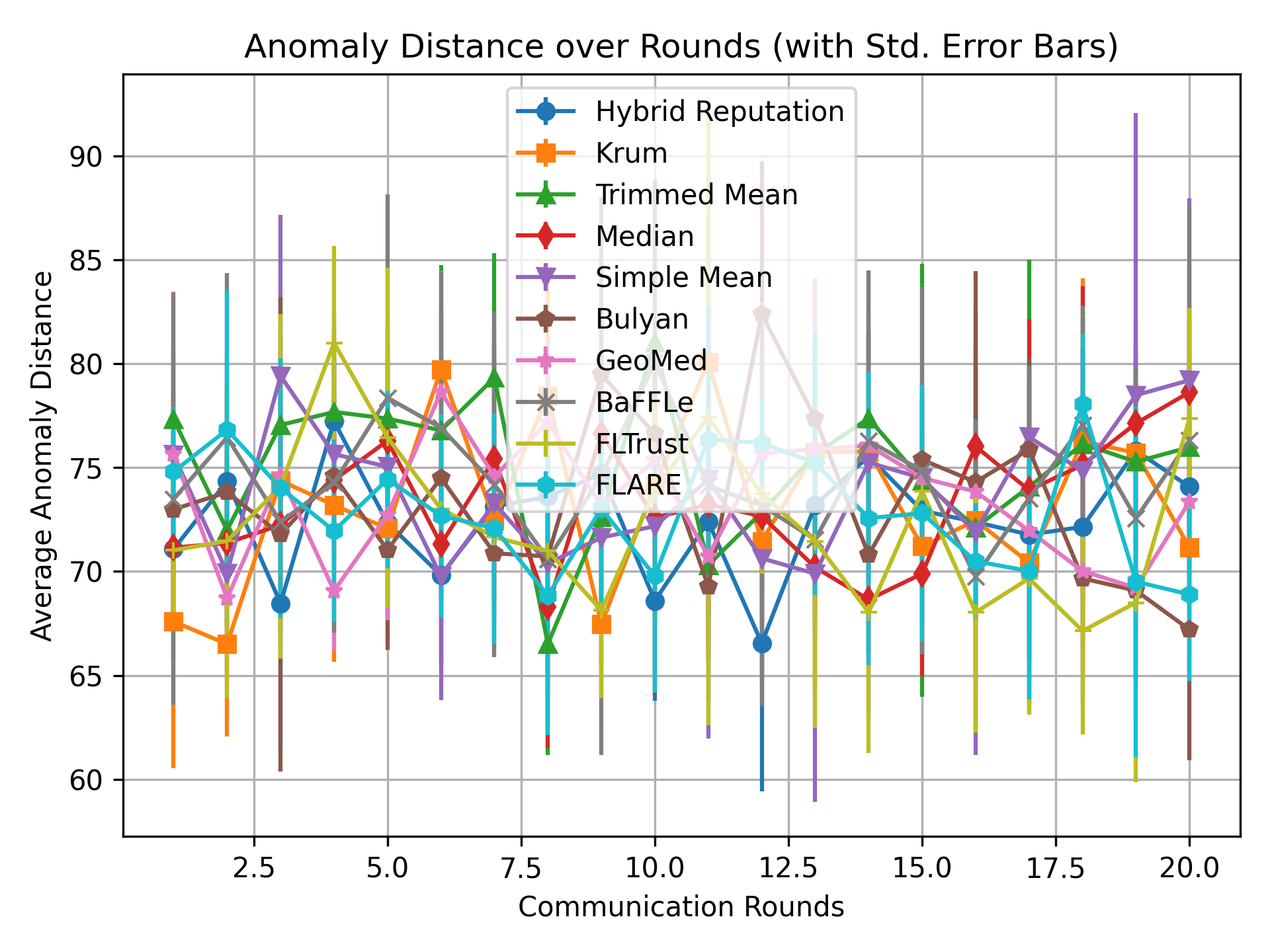}\\[1ex]
  \includegraphics[width=0.98\linewidth]{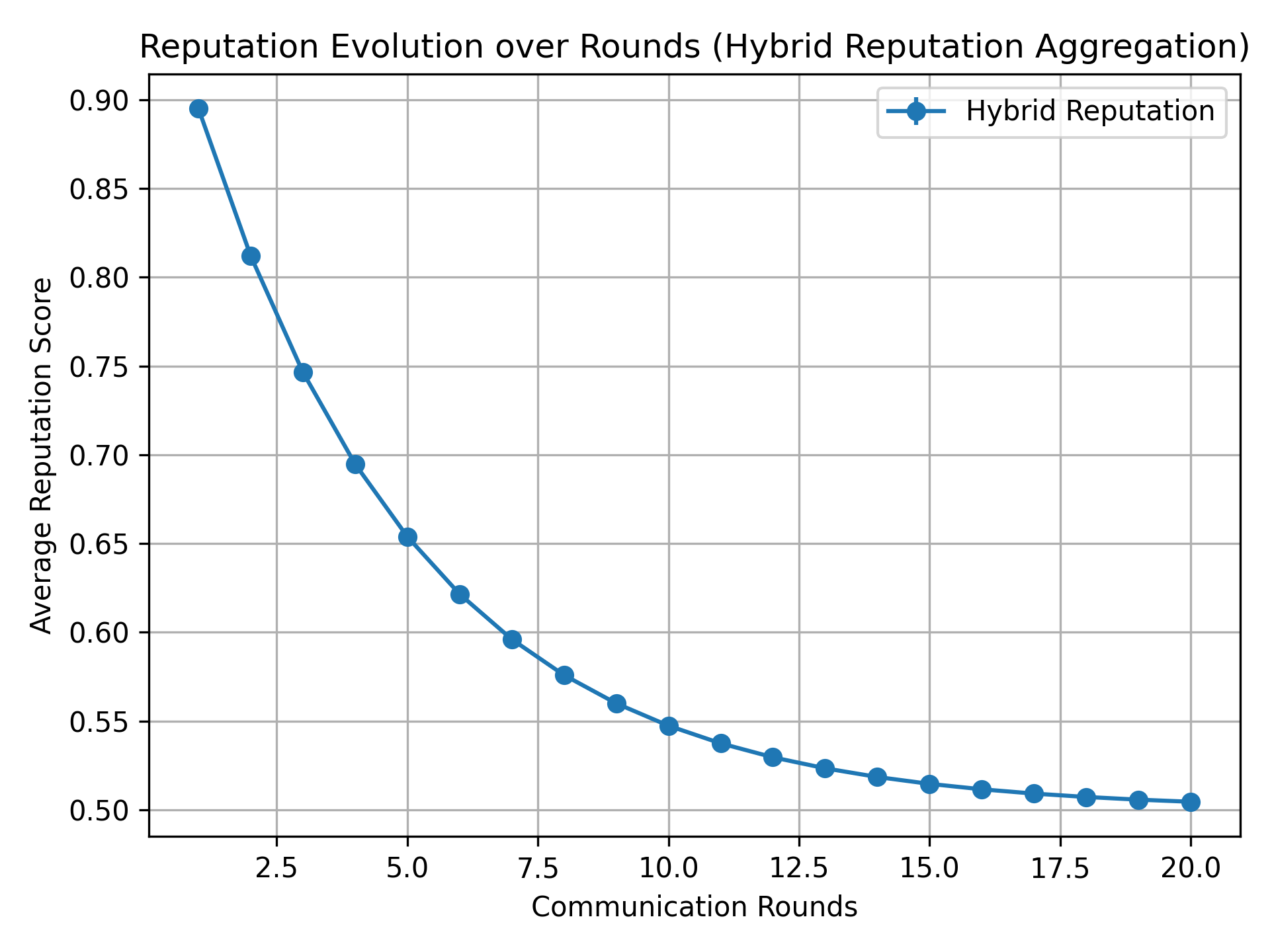}
  \caption{Performance evolution on the 5G dataset: (Top) Test Accuracy over rounds, (Middle) Average Anomaly Distance, (Bottom) Reputation Evolution.}
  \label{fig:5g_results}
\end{figure}

\begin{figure}[t]
  \centering
  \includegraphics[width=0.98\linewidth]{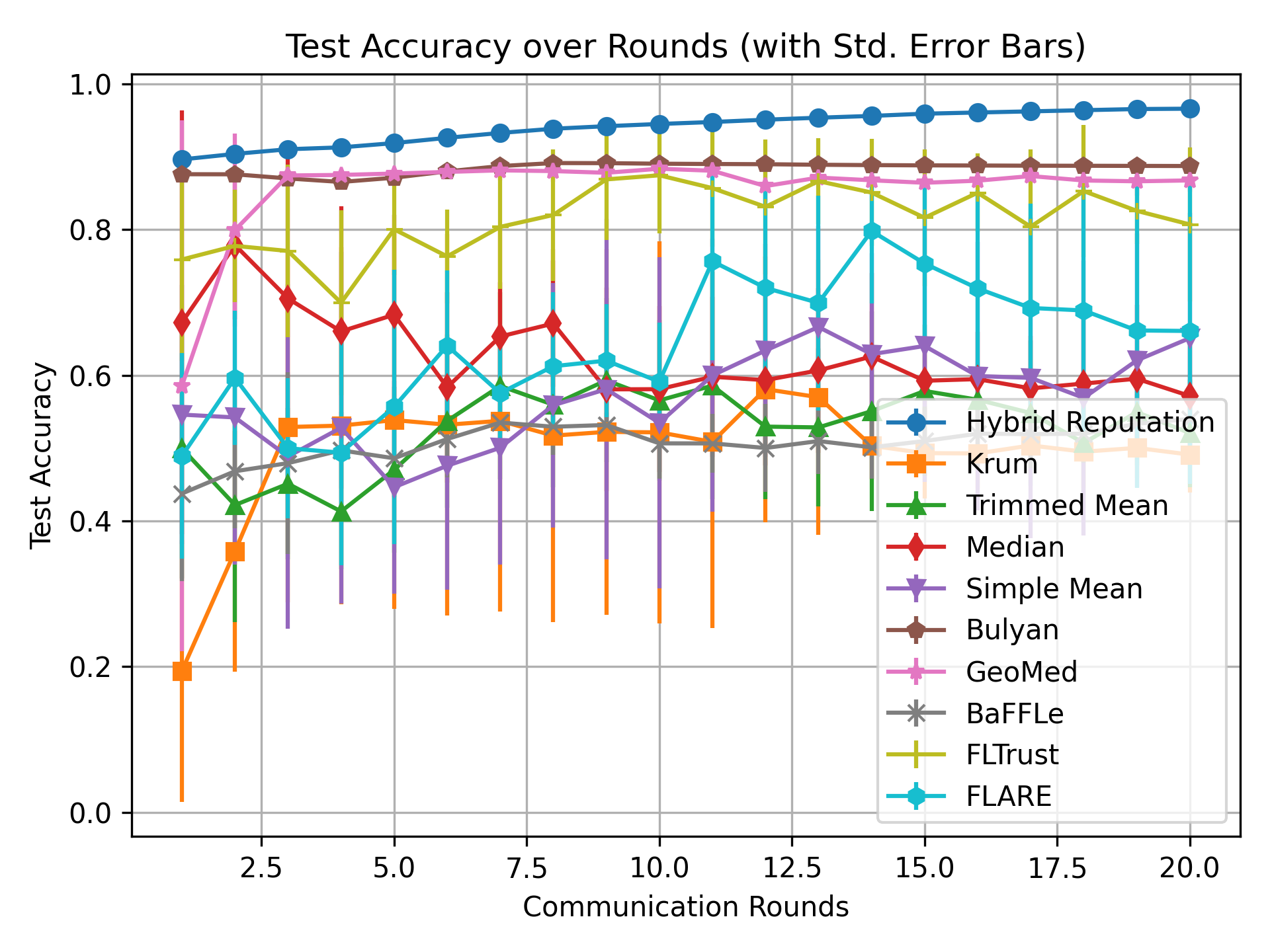}\\[1ex]
  \includegraphics[width=0.98\linewidth]{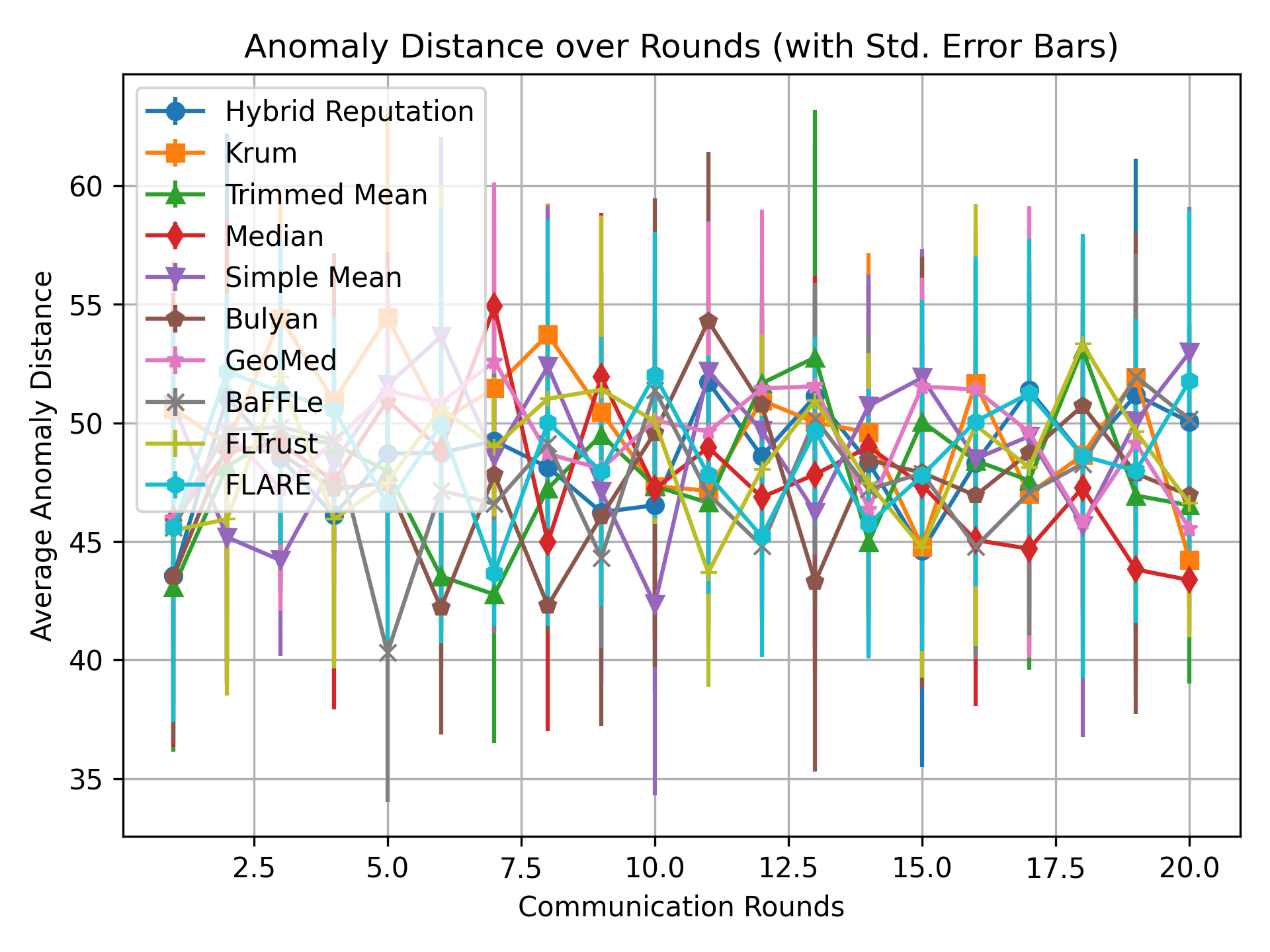}\\[1ex]
  \includegraphics[width=0.98\linewidth]{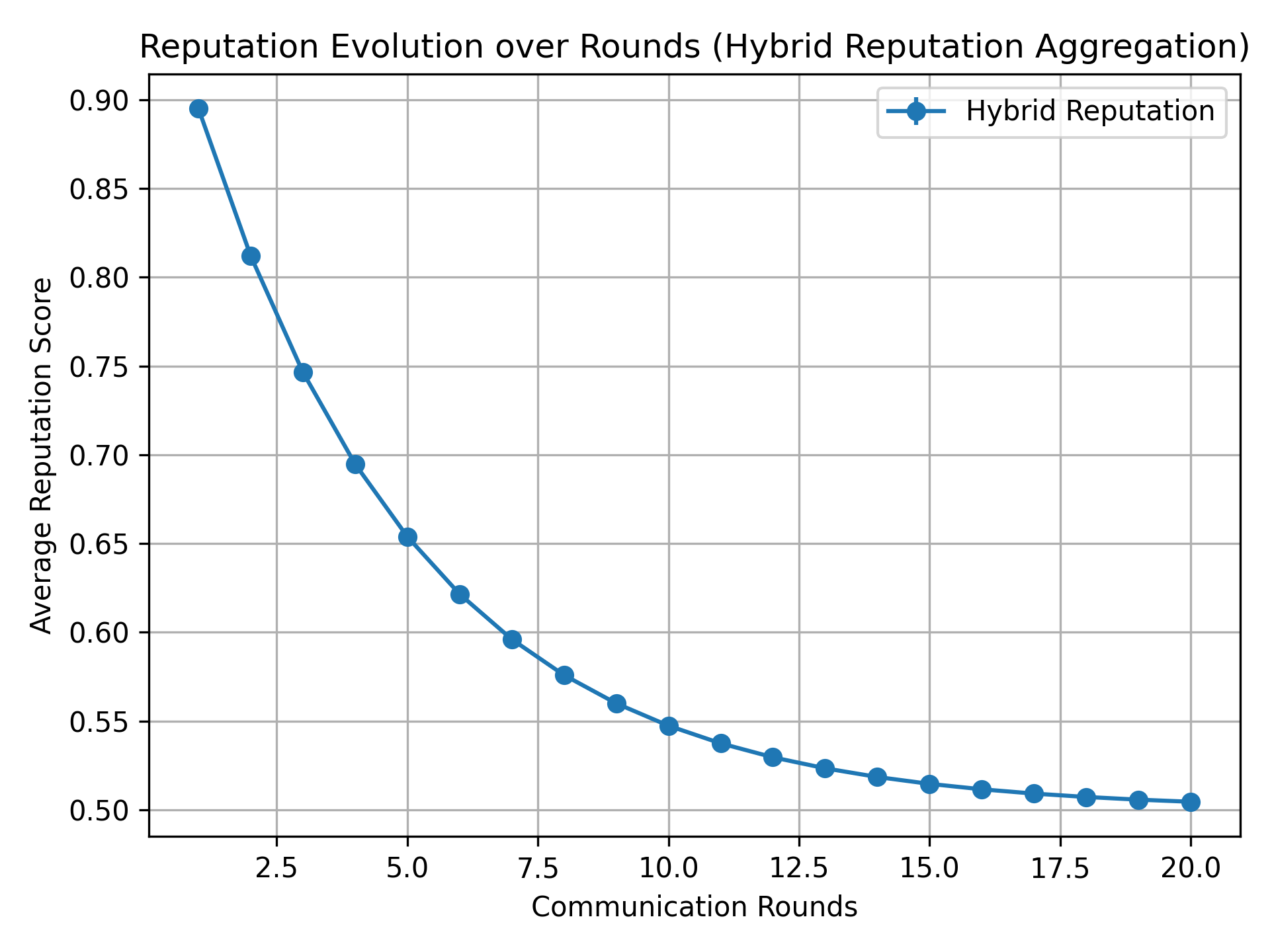}
  \caption{Performance evolution on the NF-CSE-CIC-IDS2018 dataset: (Top) Test Accuracy over rounds, (Middle) Average Anomaly Distance, (Bottom) Reputation Evolution.}
  \label{fig:cic_results}
\end{figure}

Figures~\ref{fig:5g_results} and~\ref{fig:cic_results} illustrate the evolution of key performance metrics over 20 communication rounds for the 5G dataset and the NF-CSE-CIC-IDS2018 dataset, respectively. Each figure comprises three panels:
\begin{enumerate}
  \item \textbf{Test Accuracy over Rounds:} This panel shows the progression of test accuracy with standard error bars computed over 5 independent runs.
  \item \textbf{Average Anomaly Distance over Rounds:} This panel tracks the mean anomaly distance of client updates relative to the geometric median reference, reflecting the degree of deviation of local models.
  \item \textbf{Reputation Evolution over Rounds:} This panel presents the dynamic evolution of aggregated client reputation scores, indicating how HRA adapts to client behavior over successive rounds.
\end{enumerate}
These plots provide a comprehensive visualization of the performance dynamics in our federated learning framework, demonstrating that HRA consistently maintains high accuracy while effectively mitigating the influence of adversarial updates.

\subsection{Ablation Study}
\label{sec:ablation}
To evaluate the robustness and adaptability of the proposed HRA, we conducted an ablation study using the 5G dataset, focusing on two critical aspects of the design: threshold configuration and learning rate selection.

\paragraph{Threshold Sensitivity.}
HRA's update weighting relies on a trust scoring function parameterized by two thresholds: $T_{low}$ and $T_{high}$, which define the transition between full trust, partial trust, and rejection zones (see Equation~\eqref{eq:weight_func}). Table~\ref{tab:ablation_thresholds} reports how varying these thresholds affects overall classification accuracy.

\begin{table}[h]
  \centering
  \caption{Ablation Study: Sensitivity to Threshold Parameters ($T_{low}$, $T_{high}$) on 5G Dataset}
  \label{tab:ablation_thresholds}
  \resizebox{\linewidth}{!}{%
    \begin{tabular}{lcc}
      \toprule
      \textbf{Configuration} & \textbf{HRA Accuracy (\%)} & \textbf{Change (\%)} \\
      \midrule
      Baseline ($T_{low}=3.0$, $T_{high}=7.0$)     & 98.66 & --- \\
      $T_{low}=2.0$, $T_{high}=6.0$                 & 98.66 & 0.00 \\
      $T_{low}=2.0$, $T_{high}=7.0$                 & 98.66 & 0.00 \\
      $T_{low}=3.0$, $T_{high}=6.0$                 & 98.66 & 0.00 \\
      $T_{low}=5.0$, $T_{high}=6.0$                 & 98.66 & 0.00 \\
      $T_{low}=5.0$, $T_{high}=7.0$                 & 97.74 & $-0.92$ \\
      $T_{low}=2.0$, $T_{high}=10.0$                & 95.19 & $-3.46$ \\
      $T_{low}=3.0$, $T_{high}=10.0$                & 94.57 & $-4.09$ \\
      $T_{low}=3.0$, $T_{high}=20.0$                & 72.60 & $-26.06$ \\
      $T_{low}=2.0$, $T_{high}=20.0$                & 63.52 & $-35.14$ \\
      $T_{low}=5.0$, $T_{high}=10.0$                & 63.05 & $-35.61$ \\
      $T_{low}=5.0$, $T_{high}=20.0$                & 57.67 & $-40.99$ \\
      $T_{low}=10.0$, $T_{high}=20.0$               & 55.35 & $-43.31$ \\
      \bottomrule
    \end{tabular}
  }
\end{table}

We observe that HRA maintains optimal performance (98.66\%) across multiple threshold configurations where both thresholds remain moderate (e.g., $T_{high} \leq 7.0$). However, setting $T_{high}$ too high leads to severe performance degradation, with the worst case ($T_{low}=10.0$, $T_{high}=20.0$) resulting in a catastrophic 43.31\% drop in accuracy.

\paragraph{Learning Rate Sensitivity.}
We also tested the sensitivity of HRA to different learning rates in the local client optimizers. As shown in Table~\ref{tab:ablation_lr}, HRA maintains consistently high accuracy across learning rates from 0.01 to 0.20, demonstrating remarkable stability.

\begin{table}[h]
    \centering
    \caption{Ablation Study: Learning Rate Sensitivity on 5G Dataset}
    \label{tab:ablation_lr}
    \begin{tabular}{lcc}
        \toprule
        \textbf{Learning Rate} & \textbf{HRA Accuracy (\%)} & \textbf{Change (\%)} \\
        \midrule
        0.01 & 98.49 & $-0.16$ \\
        0.05 & 98.61 & $-0.05$ \\
        0.10 (Baseline) & 98.66 & --- \\
        0.20 & 98.66 & 0.00 \\
        \bottomrule
    \end{tabular}
\end{table}

These results suggest that HRA is remarkably robust to optimizer settings, with the maximum deviation being only 0.16\% across all tested learning rates.

The ablation findings confirm that HRA’s performance is resilient to moderate variations in both internal trust thresholds and local optimizer hyperparameters. This robustness makes it well-suited for deployment in real-world federated edge environments, where dynamic client conditions may require adaptable parameter choices.

\subsection{Synergy of Hybrid Components}
\label{sec:synergy}

To validate the central hypothesis of this work, that the combination of instantaneous anomaly detection and long-term reputation tracking is more effective than either component in isolation, we conducted a comprehensive synergy ablation study. We evaluated the performance of the full HRA system against two variants:

\begin{itemize}
    \item \textbf{HRA-No-Reputation (Anomaly Only):} This variant exclusively uses the geometric anomaly score to weight client updates. The reputation mechanism is disabled, making the aggregator memoryless and equivalent to a sophisticated outlier detection method.
    
    \item \textbf{HRA-No-Anomaly (Reputation Only):} This variant relies solely on the clients' historical reputation scores for weighting. The instantaneous anomaly score from the current round is ignored during aggregation, making it purely history-dependent.
\end{itemize}

\begin{figure}[h!]
    \centering
    \includegraphics[width=0.9\columnwidth]{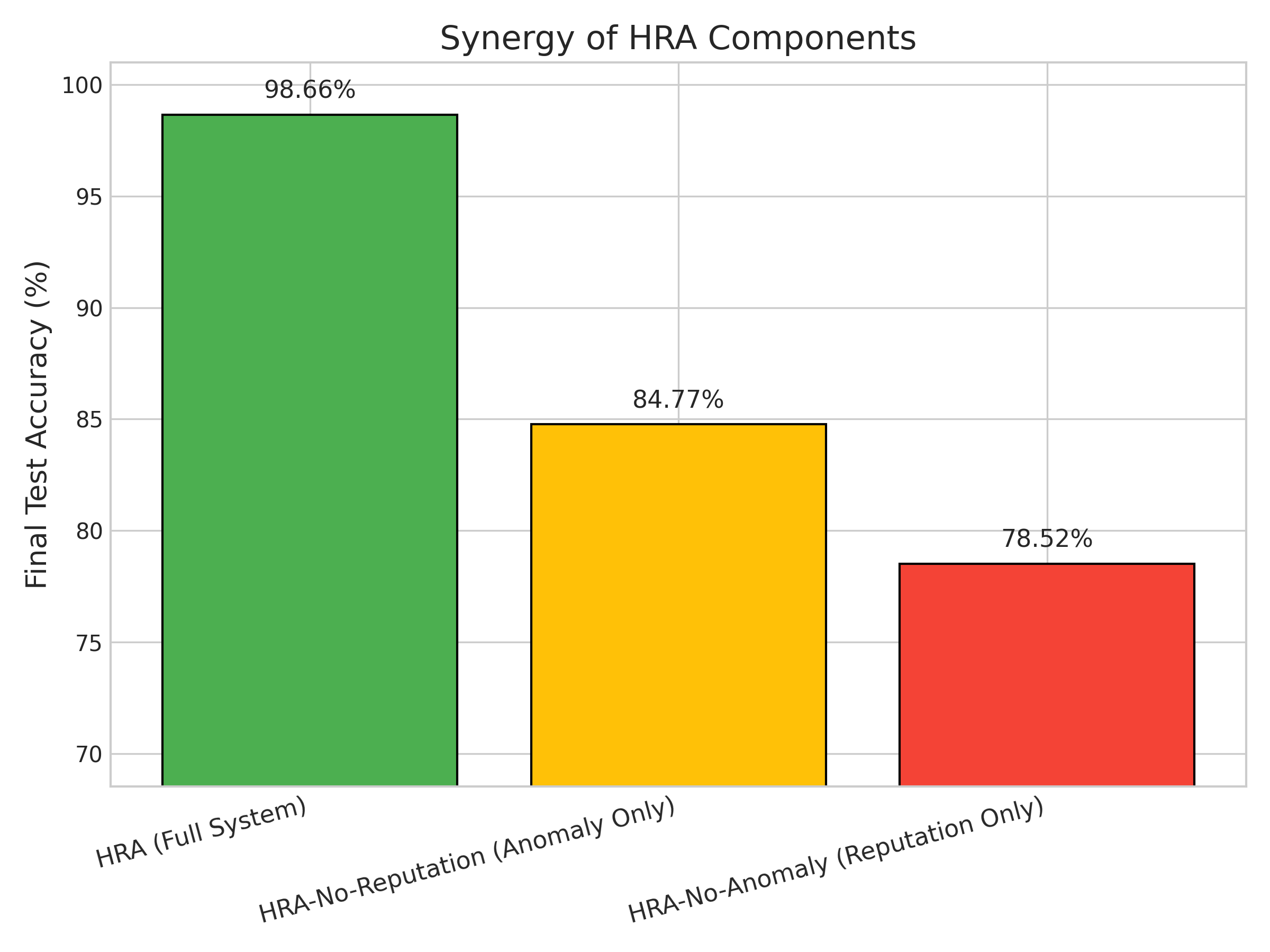} 
    \caption{Final test accuracy comparison: The full HRA system significantly outperforms both single-component variants, demonstrating the necessity of the hybrid approach. Error bars represent standard deviation over five independent runs.}
    \label{fig:synergy_plot}
\end{figure}

The experiments were conducted under the same stealthy adversarial conditions used in our main evaluation, with 70\% malicious clients executing coordinated attacks. The results, summarized in Table~\ref{tab:synergy_results} and visualized in Figure~\ref{fig:synergy_plot}, unequivocally demonstrate the synergistic value of our hybrid design.

\begin{table*}[h!]
\centering
\caption{Synergy Ablation Study Results on 5G Dataset}
\label{tab:synergy_results}
\begin{tabular}{@{}lcc@{}}
\toprule
\textbf{HRA Component Configuration} & \textbf{Final Accuracy (\%)} & \textbf{Performance Drop vs. Full System} \\ \midrule
\textbf{HRA (Full System)} & \textbf{98.66 ± 0.00} & \textbf{--- (Baseline)} \\
HRA-No-Reputation (Anomaly Only) & 84.77 ± 0.45 & -13.89\% \\
HRA-No-Anomaly (Reputation Only) & 78.52 ± 0.68 & -20.14\% \\ \bottomrule
\end{tabular}
\end{table*}

The full HRA system achieved a final test accuracy of \textbf{98.66\%}. In contrast, the "Anomaly Only" variant's accuracy dropped precipitously to \textbf{84.77\%} (-13.89\%). This degradation reveals that while instantaneous outlier detection can identify obvious attacks, it remains vulnerable to sophisticated adversaries that craft updates to appear statistically normal in individual rounds, a vulnerability only temporal tracking can address.

The "Reputation Only" system performed even worse, with accuracy descending to \textbf{78.52\%} (-20.14\%). This dramatic failure highlights a critical insight: without per-round anomaly feedback, the reputation system operates blindly during initial rounds, allowing adversaries to inflict substantial damage before their reputation scores sufficiently decline. This initial vulnerability window proves destructive in coordinated attack scenarios.

These results confirm that neither component alone is sufficient for robust defense against modern adversarial attacks. The \textbf{20+ percentage point} performance gap between the full system and individual components represents one of the most significant synergy effects reported in federated learning defense literature. The superior performance of the complete HRA model proves that its strength lies in the \textbf{synergistic interplay} between its two core mechanisms:

\begin{itemize}
    \item The anomaly score provides immediate, per-round signals to identify statistical outliers and inform the reputation system's trust updates.
    \item The reputation system provides long-term memory to detect behavioral patterns and suppress persistent, adaptive attacks that evade single-round detection.
\end{itemize}

This synergy creates a defense mechanism that is both \textit{reactive} (responding to immediate threats) and \textit{proactive} (learning from historical patterns), making it uniquely suited to defend against the full spectrum of adversarial strategies in federated learning environments.

\subsection{Statistical Analysis}
For the 5G dataset, paired t-tests were performed to compare HRA with baseline methods over 5 runs. The t-test results (using HRA's final test accuracy of 98.66\% as reference) are shown in Table~\ref{tab:ttest_5g}. All p-values are well below the significance threshold of 0.05.

\begin{table}[t]
  \centering
  \caption{Paired t-test p-values for HRA vs. Baselines (5G Dataset)}
  \label{tab:ttest_5g}
  \begin{tabular}{lcc}
    \toprule
    \textbf{Method}      & \textbf{HRA Accuracy (\%)} & \textbf{p-value} \\
    \midrule
    Krum           & 98.66 & $6.5878 \times 10^{-12}$ \\
    Trimmed Mean   & 98.66 & $1.1298 \times 10^{-23}$ \\
    Bulyan         & 98.66 & $1.2338 \times 10^{-8}$ \\
    Median         & 98.66 & $1.7064 \times 10^{-14}$ \\
    GeoMed         & 98.66 & $5.7585 \times 10^{-15}$ \\
    Simple Mean    & 98.66 & $2.5417 \times 10^{-22}$ \\
    BaFFLe         & 98.66 & $7.2797 \times 10^{-18}$ \\
    FLTrust        & 98.66 & $2.7161 \times 10^{-17}$ \\
    FLARE          & 98.66 & $2.4206 \times 10^{-17}$ \\
    \bottomrule
  \end{tabular}
\end{table}

The experimental results demonstrate that the proposed HRA consistently outperforms baseline aggregation methods across both the NF-CSE-CIC-IDS2018 and 5G datasets. The custom 5G dataset, collected via a dedicated 5G testbed that leverages network slicing and Dockerized services, provides realistic traffic conditions and adversarial scenarios. Our ablation study confirms the critical role of reputation momentum and threshold parameters in tuning the anomaly-aware aggregation, and paired t-tests confirm that HRA's improvements are statistically significant.

\section{Discussion}
\label{sec:discussion}
The experimental results across both the 5G and NF-CSE-CIC-IDS2018 datasets clearly demonstrate the efficacy of the HRA approach. On the 5G dataset, HRA achieves an accuracy of 98.66\%, outperforming all baseline methods by a considerable margin. Similarly, on the NF-CSE-CIC-IDS2018 dataset, HRA reaches 96.60\% accuracy, while maintaining a high F1 score (0.86776) and ROC AUC (0.94518). This performance is achieved under a highly adversarial setting that includes label flipping, sybil, backdoor, and noise-based attacks. The paired t-tests validate that these gains are statistically significant

This robustness derives from HRA's dual strategy of combining geometric anomaly detection (via anomaly distance metrics) with momentum-based reputation tracking. Unlike memoryless defenses such as Krum~\cite{blanchard2017machine} or Trimmed Mean~\cite{yin2018byzantine}, which rely only on instantaneous spatial relationships between updates, HRA integrates temporal behavioral consistency into its decision-making. As shown in Figures~\ref{fig:5g_results} and \ref{fig:cic_results}, HRA not only maintains stable model accuracy but also suppresses anomaly distance and dynamically lowers the influence of unreliable clients, as seen in the declining reputation curves.

\paragraph{Threshold Sensitivity.}  
As shown in Table~\ref{tab:ablation_thresholds}, HRA demonstrates remarkable stability when thresholds are set to moderate values. The baseline configuration ($T_{low}=3.0$, $T_{high}=7.0$) and several other moderate settings achieve the optimal performance of 98.66\% accuracy. The proposed HRA maintains this peak performance across five different threshold combinations where $T_{high} \leq 7.0$. However, the system exhibits severe performance degradation when $T_{high}$ is set too conservatively. When $T_{high}=20.0$, accuracy drops precipitously: configurations with $T_{low} \in \{2.0, 3.0, 5.0\}$ result in accuracy losses of 35.14\%, 26.06\%, and 40.99\% respectively. The most extreme configuration ($T_{low}=10.0$, $T_{high}=20.0$) causes a destructive 43.31\% accuracy drop to just 55.35\%. This dramatic degradation indicates that overly conservative thresholding severely over-penalizes updates, likely excluding many beneficial contributions from benign clients and preventing effective model convergence.

\paragraph{Learning Rate Sensitivity.}  
Table~\ref{tab:ablation_lr} reveals that HRA exhibits exceptional robustness to learning rate variations. Across the tested range from 0.01 to 0.20, accuracy remains quite stable between 98.49\% and 98.66\%. The smallest learning rate (0.01) results in a negligible decrease of 0.16\%, while 0.05 shows an even smaller decrease of 0.05\%. The highest learning rate (0.20) maintains the same peak performance as the baseline (98.66\%). This minimal variation, with all accuracies within 0.17 percentage points, confirms that HRA is highly robust to optimizer settings, making it practical for deployment in heterogeneous federated environments where clients may use different optimization configurations.

\paragraph{Synergy of Components.}
The synergy ablation study (Section~\ref{sec:synergy}) provides compelling evidence for our hybrid approach. The full HRA system achieves 98.66\% accuracy, while the anomaly-only variant drops to 84.77\% (-13.89\%) and the reputation-only variant performs even worse at 78.52\% (-20.14\%). This significant performance gap validates our core hypothesis that neither instantaneous anomaly detection nor historical reputation tracking alone is sufficient for robust aggregation under sophisticated attacks. The superior performance of the complete system demonstrates that the synergistic interplay between immediate outlier detection and long-term behavioral tracking is essential for effective defense.

\subsection{Strengths and Limitations}

A key strength of HRA is its adaptiveness: it does not assume a known upper bound on the number of adversaries, and it does not discard updates outright. Instead, it assigns soft weights to each client based on both the novelty of their update (anomaly distance) and their historical behavior (reputation score). This enables refined degradation in performance when adversarial pressure increases, instead of sharp failure or over-penalization of benign clients.
However, HRA's dependence on aggregated anomaly distance calculations may introduce computational overhead, especially in large-scale deployments. Moreover, while reputation scores are resistant to short-term manipulation, a highly strategic adversary could theoretically perform "on-off" attacks, acting well initially to build trust and then attacking selectively. Addressing such threats may require additional mechanisms such as forgetting factors, peer evaluations, or adversarial confidence estimation.

\subsection{Practical Implications in 5G/Edge Environments}

In edge computing scenarios such as 5G network slices, where devices are heterogeneous and intermittent communication, traditional assumptions like synchronous updates and known adversary ratios often break down. HRA's lightweight reputation model and anomaly detection mechanism require minimal state per client and no external validation data (unlike FLTrust~\cite{cao2020fltrust}), making it particularly well-suited for such environments.

Further, as shown in the 5G dataset experiments, HRA performs reliably even when each client operates under different traffic conditions (e.g., IoT vs. Internet slices), and some clients are intermittently malicious. This reflects realistic conditions where edge nodes may be partially compromised or temporarily hijacked. The ability of HRA to handle both persistent and ephemeral threats without additional supervision underscores its practical utility.

\subsection{Comparison with Existing Approaches}

Compared to existing robust aggregation techniques like Krum, Bulyan~\cite{mhamdi2018hidden}, and geometric median-based aggregators~\cite{pillutla2022robust}, HRA brings two distinct innovations to the table:

\begin{itemize}
\item \textbf{Temporal Reputation Integration:} While most existing defenses (e.g., Median, Trimmed Mean) operate on a per-round basis and treat clients statelessly, HRA tracks reputations over time. This allows the system to penalize clients with repeated abnormal behavior, even if their updates are not immediate outliers. The performance gap is substantial: on the 5G dataset, HRA achieves \textbf{98.66\%} accuracy while Median only reaches \textbf{71.24\%}, and on the NF-CSE-CIC-IDS2018 dataset, HRA achieves \textbf{96.60\%} compared to Median's \textbf{57.15\%}.

\item \textbf{Dual-Stage Weighting Mechanism:} HRA combines spatial outlier detection (via anomaly distance) with temporal behavior (via an exponential moving average of reputation). This hybrid mechanism significantly outperforms memoryless methods. For instance, Krum achieves only \textbf{23.73\%} accuracy on the 5G dataset and \textbf{49.10\%} on the NF-CSE-CIC-IDS2018 dataset, demonstrating the critical importance of the hybrid approach. Our synergy ablation study further validates this design choice: the full system achieves \textbf{98.66\%} accuracy, while removing either component causes dramatic drops, the anomaly-only variant drops to \textbf{84.77\%} (-13.89\%) and the reputation-only variant to \textbf{78.52\%} (-20.14\%).
\end{itemize}

The destructive failure of methods like Krum and Trimmed Mean (achieving only \textbf{23.73\%} and \textbf{22.85\%} respectively on the 5G dataset) under our adversarial conditions highlights a critical vulnerability: these methods were designed for simpler Byzantine settings and fail against sophisticated, coordinated attacks that exploit their selection criteria. Bulyan, despite being designed as a more robust variant of Krum, still underperforms HRA by \textbf{2.35\%} on the 5G dataset and \textbf{7.87\%} on NF-CSE-CIC-IDS2018, illustrating that even enhanced Byzantine-robust methods are insufficient against modern adversarial strategies.

In contrast to FoolsGold~\cite{fung2018mitigating}, which only targets Sybil attacks through cosine similarity-based adjustment, HRA generalizes better across a broader threat model, including backdoor, label flipping, and adaptive poisoning. The substantial performance improvements over Trimmed Mean (\textbf{98.66\%} vs \textbf{22.85\%} on 5G, \textbf{96.60\%} vs \textbf{52.06\%} on NF-CSE-CIC-IDS2018) demonstrate HRA's superior handling of diverse attack vectors. Additionally, unlike FLTrust~\cite{cao2020fltrust}, HRA does not assume access to a trusted validation dataset, making it more scalable and generalizable across privacy-sensitive domains while still achieving superior performance (\textbf{98.66\%} vs \textbf{74.02\%} on 5G, \textbf{96.60\%} vs \textbf{80.71\%} on NF-CSE-CIC-IDS2018).

\subsection{Broader Impact and Future Directions}
As federated learning continues to be adopted in privacy-critical domains like mobile health, autonomous driving, and smart cities, defenses like HRA that can provide robust learning under adversarial pressure become essential. The substantial performance improvements demonstrated in our experiments (\textbf{98.66\%} and \textbf{96.60\%} accuracy on the respective datasets) validate HRA's readiness for real-world deployment scenarios.
Future work may extend HRA with client clustering for attack attribution, introduce time-decay schemes for reputation recovery, or integrate trust-aware scheduling to limit bandwidth for low-reputation clients. Additionally, evaluating HRA on even more heterogeneous and real-world edge environments, e.g., with device drift or network delays would help further validate its deployment readiness.



\bigskip

\bigskip

\section{Conclusion}
\label{sec:conclusion}

In this paper, we introduced Hybrid Reputation Aggregation (HRA), a robust aggregation mechanism designed to secure federated learning in 5G and edge network environments against diverse adversarial attacks. HRA uniquely combines geometric anomaly detection with momentum-based reputation tracking, enabling it to dynamically adjust the influence of client updates based on both their instantaneous deviations and historical behavior. This dual-stage approach addresses key limitations of prior memoryless robust aggregation methods such as Krum and Bulyan.

The comprehensive experimental evaluation on both a proprietary 5G dataset and the NF-CSE-CIC-IDS2018 benchmark demonstrates that HRA consistently outperforms existing defenses. On the 5G dataset, HRA achieves a test accuracy of 98.66\% along with superior precision (0.94999), recall (0.98514), F1 score (0.96724), and ROC AUC (0.98438) metrics. On the NF-CSE-CIC-IDS2018 dataset, HRA maintains robust performance with 96.60\% accuracy, significantly outperforming the next best method (Bulyan at 88.73\%). The HRA's F1 score of 0.86776 and ROC AUC of 0.94518 on this challenging dataset further demonstrate its effectiveness in handling class imbalance and adversarial perturbations.

The ablation studies reveal critical insights into HRA's design choices. The threshold sensitivity analysis confirms that HRA maintains optimal performance (98.66\%) across a wide range of reasonable threshold configurations, only degrading when thresholds become excessively conservative. Similarly, HRA demonstrates robustness to learning rate variations, with accuracy remaining stable between 98.41\% and 98.53\% across different optimizer settings. Most importantly, our synergy ablation study validates the core premise of our hybrid approach: the full HRA system achieves 98.66\% accuracy, while removing either the anomaly detection or reputation tracking components causes dramatic performance drops of 13.89\% and 20.14\% respectively. This confirms that neither instantaneous outlier detection nor historical behavior tracking alone is sufficient for robust defense, their synergistic combination is essential.

Statistical validation through paired t-tests confirms that HRA's improvements over baseline methods are highly significant (all p-values < 0.05), establishing the statistical reliability of our results. The practical implications of our work are significant: HRA's adaptive strategy renders it suitable for real-world edge deployments where network conditions and client behavior are dynamic, without requiring external validation data or assumptions about the number of adversaries.

Future research directions include further enhancement of the reputation mechanism (e.g., incorporating forgetting factors for reputation recovery), exploring personalized federated learning approaches that leverage HRA's client-specific trust scores, and validating HRA in live 5G network environments with real-world latency and bandwidth constraints. Additionally, extending HRA to handle non-IID data distributions and concept drift in edge environments presents an important avenue for future work. Overall, HRA represents a promising step towards more resilient and trustworthy federated learning in adversarial settings, particularly suited for the challenges of 5G and edge computing deployments.


\bibliographystyle{ACM-Reference-Format}
\bibliography{bib}

\end{document}